\documentclass{article}
\usepackage{bigsky2007}
\usepackage{graphicx}
\frompage{000} \topage{000}                                              
\def\rightmark{Particle Production from RHIC to LHC}
\def\leftmark{R. Bellwied}

\title{Particle production mechanisms from RHIC to LHC}
\authors{
{Rene Bellwied$^1$ %
}\\[2.812mm]
{\normalsize
\hspace*{-8pt}$^1$ Wayne State University, Physics Department\\
666 West Hancock, Detroit, MI 48201, U.S.A.\\
e-mail: bellwied@physics.wayne.edu\\[0.2ex]
}}

\abstract{I will review RHIC data with respect to the intriguing
possibility that the hadron production mechanism in the produced
partonic medium might be different than in vacuum. I will use the
measurements of collective features, such as flow and quenching of
identified particles, to show that different regions of the particle
momentum spectrum are likely populated through different mechanisms,
and that the medium seems to play an important role in
hadronization. Finally I will address the question whether the
different initial conditions achievable in heavy ion collisions at
LHC energies, compared to RHIC, might affect the properties of the
deconfined quark-gluon phase and its hadronization to baryonic
matter.}

\PACS{specifications see, e.g.\ {\tt http://www.aip.org/pacs/}}

\begin{document}

\maketitle
\setcounter{page}{1}

\section{Introduction}\label{intro}

It has long been assumed that the hadronization or fragmentation
process in vacuum can be described by string models, which
distinguish between longitudinal string fragmentation to explain the
soft underlying event and transverse parton fragmentation to explain
the hard scattering jet formation mechanism \cite{feynman}. Although
no explicit hadronization mechanism is given in this picture the
particle formation can be parametrized through the fragmentation
function D$_{q}^{h}$, which yields the probability that a certain
parton 'q' fragments into a certain hadron 'h'. Baryon formation is
especially difficult to visualize in such a model and requires in
most approaches the formation of a di-quark structure, as a remnant
of the initial hard scattering in a proton-proton collision. Recent
results from RHIC have actually contributed significantly to the
more detailed understanding of hadronization in vacuum. I will
discuss these results briefly in the next chapter.

Hadronization in medium should be different simply because the
majority of partons which contribute to the hadron formation are
likely to equilibrate to a thermal state before hadronization
occurs. In that sense we do not expect to see features of a.) jet
formation and b.) string fragmentation at least in the low momentum
part of the particle emission spectra. Early RHIC measurements have
shown though that we can expect to see the characteristics of
modified (due to quenching) jet formation at sufficiently high
transverse momenta (p$_{T}$). In addition, the intermediate p$_{T}$
range shows features of a recombination mechanism, using either
thermal \cite{bass} or non-thermal \cite{hwa} partons in an
additional hadronization process. Differential measurements of the
flavor dependence of collective phenomena such as elliptic flow and
jet quenching should allow us to better understand hadron formation.
I will review the details of the first five years of identified
particle measurements at RHIC and show that the measured flavor
dependencies are not well understood and lead to new questions about
hadron production in medium.

In the final chapter I will briefly review the anticipated changes
in the initial conditions of the partonic phase when the collision
energy is increased from RHIC to LHC. In particular I will argue
that the strong coupling seen in the Quark Gluon Liquid at RHIC is
likely to reduce to a point where the early phase is a weakly
interacting plasma. I will comment on the effects that this drastic
change in conditions might have on the hadronization and the
collective phenomena measured at RHIC.

\section{Hadronization in vacuum}\label{vacuum}

Over the past five years the RHIC experiments have studied particle
formation in proton-proton collisions in great detail. These studies
go beyond the initial ISR studies \cite{isr} and even the recent
FNAL studies \cite{fnal}, in terms of particle identification
capabilities and the application of modern analysis methods, which
are largely based on the analysis of heavy ion reactions.  The most
relevant results regarding hadronization out of the vacuum are:

a.) breakdown of the so-called m$_{T}$-scaling at intermediate
p$_{T}$ as shown in Fig.1 \cite{star-str}. Instead of a common
scaling for all identified m$_{T}$ spectra, STAR has found a
baryon/meson scaling at sufficiently high transverse momentum. This
can be explained by the requirement of di-quark formation for baryon
production, which leads to a di-quark suppression factor which needs
to be applied to the baryon spectra in order to find a common hadron
scaling. This effect is well described by the gluon fragmentation
model in PYTHIA. It is the first experimental evidence for di-quark
formation at RHIC though.

\begin{figure}
\hspace{2.5cm}
\includegraphics[width=2.9in, bb=0 0 800 800]{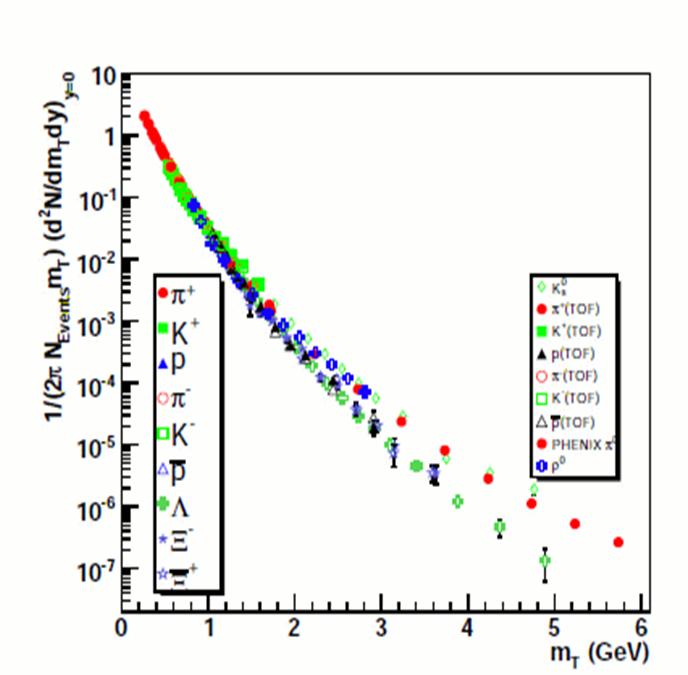} \label{fig:1}
\caption{Scaled transvere mass mid-rapidity (y = +-0.5) spectra
measured in 200 GeV proton proton collisions in STAR and PHENIX
\cite{star-str}. }
\end{figure}

b.) gluon dominance in the fragmentation process at RHIC energies.
Besides the m$_{T}$ scaling, the lack of discernable differences in
the particle vs. anti-particle production over the kinematic range
measured at RHIC, and the enhanced gluon fragmentation contribution
in PYTHIA and fragmentation function fits \cite{akk}, necessary to
describe RHIC data \cite{star-str,star-spec}, shows that at these
collision energies the parton interactions are indeed dominated by
low x gluons.

c.) contributions of non-valence quark fragmentation to, in
particular, baryon production at RHIC. This effect is best
documented by several new parametrizations of the baryon
fragmentation functions by Albino et al. \cite{akk}, Bourelly and
Soffer \cite{bs}, and DeFlorian et al. \cite{def}.

Two particle correlation measurements of unidentified charged
particles in pp and AA collisions have also contributed
significantly to the understanding of the hadronization process in
vacuum and in medium. In vacuum the non-triggered correlation
functions can be separated into a soft longitudinal string
fragmentation which generates the 'underlying event', and a hard
transverse parton fragmentation process which generates the jet
structures \cite{austin}. This differential measure develops into a
medium dependent pattern as a function of centrality in AA
collisions. A complementary triggered analysis based on the leading
particle momentum in pp and AA collisions shows that the jet
structure itself, as determined by its charge ordering, shows very
little centrality dependence as long as the jets are emitted near
the surface of the fireball \cite{lll,putschke}. This has been
confirmed recently in identified particle correlations measurements
\cite{bielcikova}.

\section{Hadronization in medium}\label{medium}

The debate about a different hadronization mechanism from a
deconfined partonic phase compared to the vacuum has been fueled by
detailed measurements of dynamic effects in identified particle
production at RHIC. The results for collective elliptic flow (v2)
and nuclear suppression (R$_{AA}$) in the intermediate p$_{T}$ range
can both be parametrized with the number of constituent quarks in
baryons and mesons, respectively. The scaled measurements are shown
in Figs. 2 and 3, respectively.

\begin{figure}
\hspace{3.cm}
\includegraphics[width=2.6in, bb=0 0 800 800]{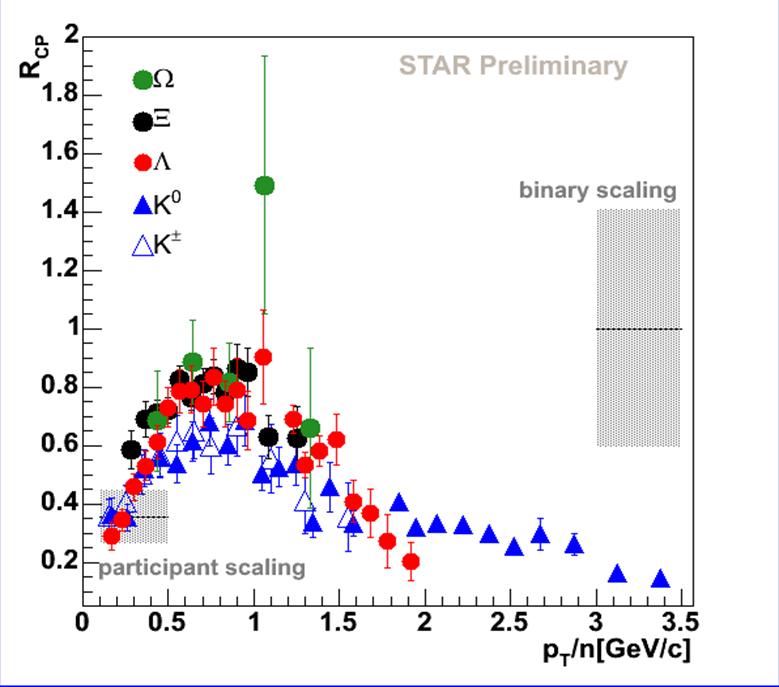} \label{fig:2}
\caption{STAR data on R$_{CP}$ vs p$_{T}$/n  in Au+Au collisions at
$\sqrt{s_{NN}}$ = 200 GeV using n=3 for baryons and n=2 for mesons.
R$_{CP}$ is calculated from 0-5\% and 40-60\% central Au+Au
collisions. }
\end{figure}

\begin{figure}
\hspace{2.cm}
\includegraphics[width=2.2in, bb=0 0 800 800]{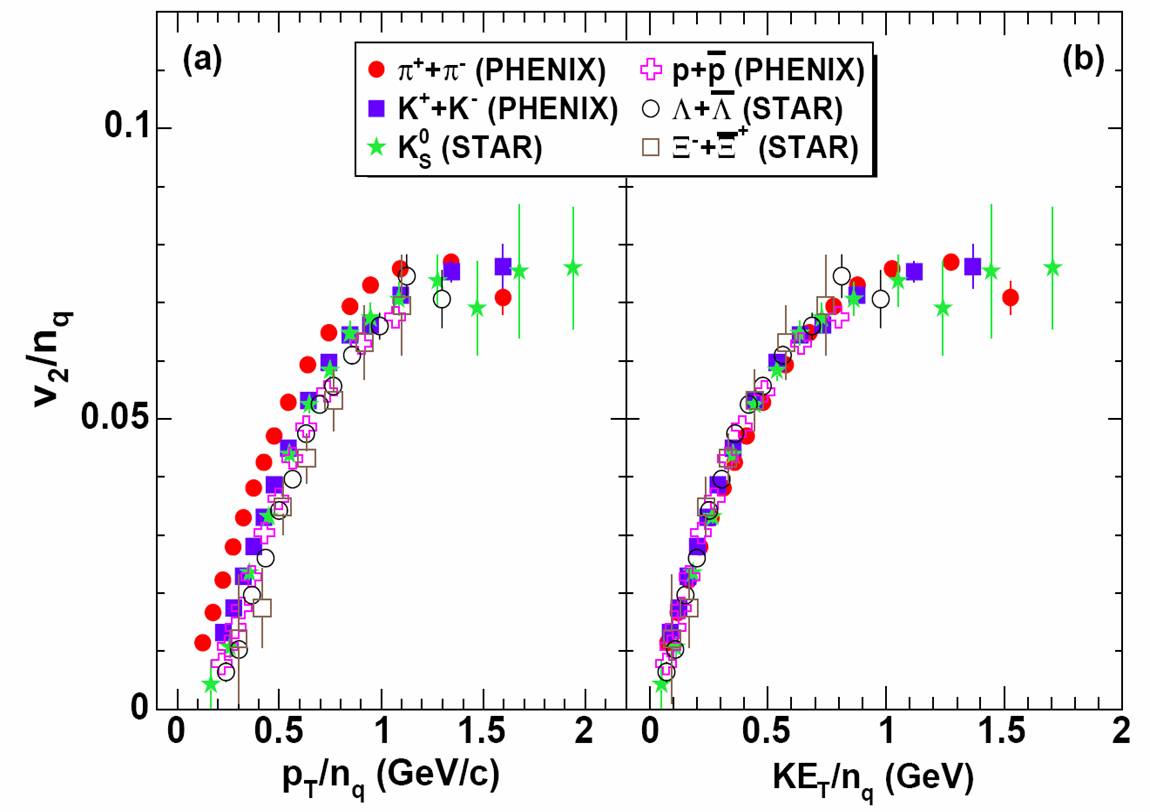} \label{fig:3}
\caption{ v2/n$_{q}$ vs. p$_{T}$/n$_{q}$ and KE$_{T}$/n$_{q}$ for
several particle species measured by STAR and PHENIX as indicated in
Au+Au collisions at $\sqrt{s_{NN}}$ = 200 GeV \cite{v2-summary}.}
\end{figure}

This gave rise to a whole series of theory papers describing the
process of partonic recombination as the main production mechanism
for hadrons from the medium. The bulk matter, which takes on the
role of the underlying event in heavy ion collisions exhibits
features of thermal emission, but unfortunately the possible
contribution of an interacting hadronic phase to the thermal
properties (radial flow, mass scaling etc.) make the interpretation
of the hadronization mechanism for low momentum particles quite
ambiguous. In the intermediate to high momentum range, where the
mass of the relevant degrees of freedom is negligible, we should be
able to probe the mechanism in a more detailed way.

In that context it is interesting to note that there is a total lack
of constituent quark mass dependence in the scaling of v2. In
recombination approaches this is largely attributed to the fact that
the constituent quark mass of the up, down and strange quarks is
quite similar (300 and 460 MeV, respectively) and that all
identified particles measured until recently did not include heavier
flavors. The recent measurement of the nuclear suppression factor
and the elliptic flow for D-mesons, based on electrons from the
semi-leptonic decay of the heavy mesons
\cite{heavy-data1,heavy-data2,heavy-data3}, should allow us to
determine the applicability of partonic recombination a little
better. In other words a heavy constituent quark should change the
pattern of the v2 and R$_{AA}$ measurements. Early results, though,
seem to indicate that this is not the case. Both, the R$_{AA}$ and
the v2 measurements, can only be explained if one assumes identical
p$_{T}$-dependencies for the flow and the quenching of light and
heavy quarks.

Many models have been proposed to address these measurements and in
particular the apparent lack of a dead cone effect for induced gluon
radiation, as well as the lack of a heavy quark mass dependence in
the v2. All these models try to give the heavy quark a special
status, by postulating either the survival of heavy quark resonant
states above T$_{c}$ \cite{rapp,rapp2} or the reduced formation time
of heavy quark hadrons from the partonic phase \cite{vitev}. Fig.4
shows a comparison of the data to the heavy quark bound state model.

\begin{figure}
\includegraphics[width=2.0in, bb=0 0 800 800]{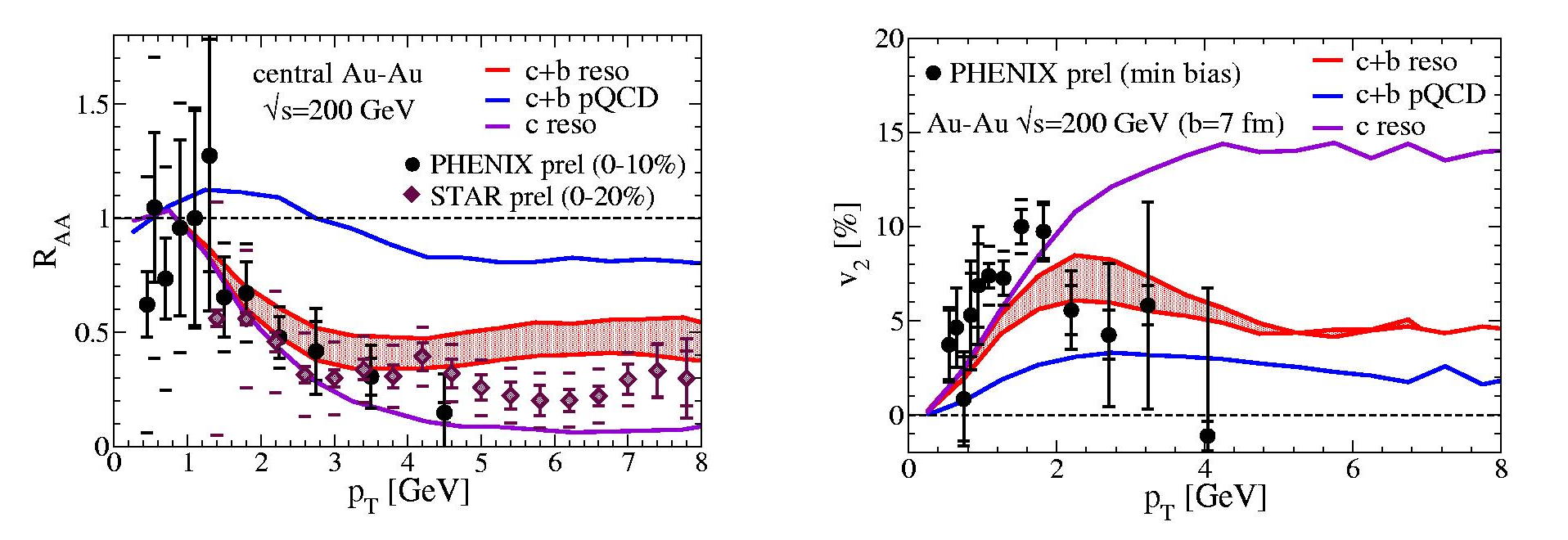} \label{fig:4}
\caption{Nuclear suppression factor (left panel) and elliptic flow
(right panel) for non-photonic single electron spectra in
semi-central Au+Au collisions at RHIC. Data
\cite{heavy-data1,heavy-data2,heavy-data3} are compared to theory
predictions \cite{rapp2} using Langevin simulations with elastic c-
and b-quark interactions in an expanding QGP fireball and
heavy-light quark coalescence at hadronization \cite{rapp}.}
\end{figure}

The near identical p$_{T}$-dependence of the v2 and the quark energy
loss between light and heavy quarks is very striking, though, and
might require a much more fundamental explanation.

One possibility is that the quasi-particle state formed near T$_{c}$
is really not depending on the constituent or even bare quark mass
concept, but rather simply the number of partons, which could be
mostly gluons, until close to hadronization. Still, for a dynamic
evolution measure such as the v2 as a function of p$_{T}$, the
dynamics of the degree of freedom has to play a role, and at least
the effect of the bare quark mass should be measurable if we indeed
probe the fragmentation or recombination of quarks. A detailed
measurement of reconstructed D-mesons and B-mesons is sorely needed
to remove the ambiguities in the semi-leptonic measurements, and a
future measurement of high momentum heavy flavor mesons and baryons
should answer the question of the applicability of recombination as
a hadronization mechanism.

Detailed measurements of strange particle production and
correlations in jets have already shown deviations from simple
recombination predictions \cite{abelev}. On the other hand medium
response effects, such as the formation of emission structures
(ridge, cone) in the wake of a traversing jet seem to again point at
different particle production mechanisms in the jet and in the
medium response structure. In the context it is interesting to note
the emission pattern differences between the medium response to the
triggered jet (same-side) and the non-triggered jet (away-side).
Although the same-side jet is expected to show a large surface bias
(hard scattering occurs near the surface of the fireball, so that
the same-side jet can escape without being quenched), the medium
traversed must be finite because the ensuing ridge structure in
$\Delta\eta$ is always correlated with a jet. On the away side the
medium response apparently leads to a cone structure in $\eta$ and
$\phi$ \cite{jia}. Early studies of the particle composition in the
ridge \cite{bielcikova} and in the cone \cite{zuo} were shown at
this conferences, and show distinct differences to jet fragmentation
in vacuum. Within the measured p$_{T}$ range the baryon to meson
ratios in the cone and the ridge actually agree with each other and
with predictions from recombination models.

\section{From RHIC to LHC}\label{energy}

Recent lattice QCD calculations, which were summarized by Peter
Petreczky at this conference \cite{petreczky} show convincingly that
the finer lattice grid and recent improvements in the staggering
algorithms lead to a quantitatively different picture than previous
calculations \cite{karsch}.

In particular, the hadronization temperature (T$_{crit}$ in lattice
QCD) is now about 20 MeV higher ($~$190 MeV) than previously, which
leads to a de-coupling of the hadronization and chemical
equilibration ($~$170 MeV) surfaces. Even more importantly the
strong coupling strength reaches the weak limit at around 3 T$_{c}$.
This fact is nicely documented by the quantitative agreement between
lattice QCD, hard thermal loop, and resummed perturbation
calculations above 3 T$_{c}$ \cite{iancu}. So it is very likely that
at LHC energies we will indeed reach the plasma, rather than liquid,
phase which was originally anticipated for RHIC energies. This phase
will only exist for a very short time (a few fm/c) and then has to
de-excite through the strong coupling phase to the hadronization
surface, but the question arises whether the weak coupling in the
early phase might lead to any measurable features. It is likely that
the hadronization mechanism is not affected, but collective
phenomena which are supposed to develop early, such as collective
elliptic flow, might be reduced by the weak coupling phase. One can
also speculate that the system might be more dilute when it enters
the strong coupling regime, and therefore exhibits less of a
collectivity.

On the other hand, the partonic system is expected to live longer,
and estimates by Eskola et al. \cite{eskola} show that the
applicability of hydrodynamics might extend to higher p$_{T}$ which
means that the thermal bulk particle formation mechanism will start
to populate the intermediate p$_{T}$ range. At the same time it is
likely that recombination will push out to higher p$_{T}$ simply
because the thermal partons will carry more energy at LHC energies
\cite{bass2}. Finally the increase in jet cross section at LHC
energies will also affect the single particle spectra in a
measurable way. Jet quenching is expected to lead to enhanced
particle production in the intermediate p$_{T}$ range
\cite{borghini}. This effect is not different from quenching at the
lower energies, but the enhanced jet cross section and the enhanced
average jet energy at the LHC contributes close to 50\% of the
particle yield in the p$_{T}$ range between 1 and 6 GeV/c. In the
model this enhancement is due to hadronization of gluon radiation,
and it still needs to be shown whether recombination can be applied
to describe soft gluon fragmentation. Certainly hybrid models which
allow the recombination of thermal partons with hard fragmentation
partons claim to predict the particle spectrum at the LHC over a
wide momentum range (2-20 GeV/c) \cite{hwa2}. It will be important
to distinguish different parton hadronization mechanisms in the mid
to high p$_{T}$ range at the LHC.

\vfill\eject
\end{document}